\documentclass[aps,prl,twocolumn,superscriptaddress]{revtex4-1}

\usepackage[pdftex]{graphicx}

\begin{document}

\title{Upper critical field of isoelectron substituted SrFe$_2$(As$_{1-x}$P$_x$)$_2$ }

\author{S.~Yeninas}
\affiliation{The Ames Laboratory, Ames, IA 50011, USA}
\affiliation{Department of Physics and Astronomy, Iowa State University, Ames, IA 50011, USA}

\author{M.~A.~Tanatar}
\affiliation{The Ames Laboratory, Ames, IA 50011, USA}
\affiliation{Department of Physics and Astronomy, Iowa State University, Ames, IA 50011, USA}

\author{J.~Murphy}
\affiliation{The Ames Laboratory, Ames, IA 50011, USA}
\affiliation{Department of Physics and Astronomy, Iowa State University, Ames, IA 50011, USA}

\author{C.~P.~Strehlow}
\affiliation{The Ames Laboratory, Ames, IA 50011, USA}
\affiliation{Department of Physics and Astronomy, Iowa State University, Ames, IA 50011, USA}

\author{O.E.~Ayala-Valenzuela}
\affiliation{NHMFL, Los Alamos National Laboratory, Los Alamos, NM 87545, USA}

\author{R.D.~McDonald}
\affiliation{NHMFL, Los Alamos National Laboratory, Los Alamos, NM 87545, USA}

\author{U.~Welp}
\affiliation{Materials Science Division, Argonne National Laboratory, Argonne, Illinois 60439, USA}

\author{W.~K.~Kwok}
\affiliation{Materials Science Division, Argonne National Laboratory, Argonne, Illinois 60439, USA}

\author{T.~Kobayashi}
\affiliation{Department of Physics, Graduate School of Science, Osaka University, Osaka 560-0043, Japan}

\author{S.~Miyasaka}
\affiliation{Department of Physics, Graduate School of Science, Osaka University, Osaka 560-0043, Japan}
\affiliation{JST, Transformative Research-Project on Iron Pnictides (TRIP), Tokyo 102-0075, Japan}

\author{S.~Tajima} 
\affiliation{Department of Physics, Graduate School of Science, Osaka University, Osaka 560-0043, Japan}
\affiliation{JST, Transformative Research-Project on Iron Pnictides (TRIP), Tokyo 102-0075, Japan}

\author{R.~Prozorov}
\email[Corresponding author: ]{prozorov@ameslab.gov}
\affiliation{The Ames Laboratory, Ames, IA 50011, USA}
\affiliation{Department of Physics and Astronomy, Iowa State University, Ames, IA 50011, USA}

\date{25 January 2013}

\begin{abstract}
The upper critical field $H_{c2}$ of optimally doped  iron-based superconductor SrFe$_{2}$(As$_{1-x}$P$_{x}$)$_{2}$ ($x$ = 0.35, $T_c$ = 25~K) was measured as a function of temperature down to 1.6~K for two principal directions of magnetic field $H \parallel c$ and $H \parallel a$. Measurements were performed in pulsed magnetic fields up to 65 T using a tunnel-diode resonator technique on as-grown and heavy-ion irradiated single crystals, with columnar defect density corresponding to a matching field $B\phi$ = 25 T. The $H_{c2,c}(T)$ is close to $T$-linear, while clear saturation is observed for $H_{c2,a}(T)$, leading to a strongly temperature dependent anisotropy parameter $\gamma$. The linear shape of $H_{c2,c}(T)$ is very similar to that observed in nodal KFe$_2$As$_2$ but very different from full-gap LiFeAs. Irradiation does not introduce any additional features on $H_{c2}(T)$ line corresponding to the matching field. Instead, it suppresses uniformly both $T_c$ and $H_{c2}$, keeping their ratio constant. 

\end{abstract}



\maketitle

\section{Introduction}

Superconductivity in the $AE$Fe$_2$As$_2$ ($AE$ = Ba, Sr or Ca) family of compounds, frequently referred to as 122 iron pnictides, can be induced in a variety of ways. It can be achieved by hole-doping with alkali $A$ metal substitution of alkali earth metal $AE$ \cite{Rotter,SrK,CaNa} as in (Ba$_{1-x}$K$_x$)Fe$_2$As$_2$, transition metal substitution on Fe site as in $AE$(Fe$_{1-x}TM_x$)$_{2}$As$_{2}$  \cite{Sefat,CaCo}, or isoelectron substitution on As site as in $AE$Fe$_2$(As$_{1-x}$P$_x$)$_2$ \cite{BaP122,SrP122,CaP122}. In the rest of the text we will label materials by the type of $AE$ and dopant elements, such as BaK122, Ba$TM$122, or BaP122. 

Approximately the same maximum $T_c$ of about 30~K is achieved for most alkali earth elements and substitutions, which suggests common origin of superconductivity for this class \cite{Paglione,Johnston,StewartRev}. Another unifying feature is universal observation of superconductivity in close proximity to stripe-type antiferromagnetic order, with maximum $T_c(x)$ observed at a doping level close to a critical point where the antiferromagnetic ordering temperature $T_N(x) \to 0$. 

Despite this common phenomenology, studies of the normal state resistivity \cite{Taillefer,BaPinterplane,TanatarNaFeAs,TanatarNaFeAs2,FisherScience} and superconducting state properties, see for example \cite{Chubukov,ProzorovreviewPD} for review, revealed unexpected diversity. For example, studies of the superconducting gap structure from penetration depth \cite{GordonPRB,MartinNi,ProzorovreviewPD}, directional heat transport \cite{TanatarPRL,Reidcaxis}, and heat capacity \cite{BNC,Ronning,Stewart,Tanigaki} as a function of doping suggested full gap at optimal doping universally in Ba$TM$122  and BaK122 \cite{ReidBaK}, evolving towards strongly anisotropic and nodal in the under-doped and overdoped regimes. A similar trend is found in the other families of iron pnictides, e.g. Co- and environmentally doped NaFeAs \cite{KyuilNaFeAs} and Ca10-3-8 \cite{Kyuil1038}. In sharp contrast, the gap is nodal in BaP122 at optimal doping \cite{Kasahara}, it remains nodal for all compositions, and clear signatures of a quantum critical point can be tracked in the properties of the superconducting condensate \cite{BaPScience}. 

Limited data about the superconducting gap structure available for compounds with other alkali earth metals, different from Ba, suggest that these unique properties of phosphorus doped compositions may be universal, and are at least observed in SrP122 \cite{SrPNMR,SrPmicrowave,SrPTDR}. It is therefore of interest to get a broader insight into the properties of P-doped materials. 

The upper critical field of iron pnictides was studied systematically over the phase diagram of BaCo122 \cite{NiNiCo,Kano} and revealed a clear distinction between underdoped and over-doped regimes. This distinction was suggested to be linked with change of the topology of the Fermi surface, however, it can be related to the evolution of the superconducting gap anisotropy in this family as well. Indeed, in stoichiometric iron-pnictide superconductors, full gap LiFeAs \cite{HKimLiFeAs,TanatarLiFeAs}, and nodal KFe$_2$As$_2$ \cite{FukazawaNMRK,ShiyanK,HashimotoK,StewartHCK,ReidK,MatsudaKScienceARPES}, studies of the upper critical fields \cite{KyuilHc2LiFeAs,LiFeAsTerashima,Balakirev,K122Terashima} reveal significantly different temperature dependences. Relatively little is known about $H_{c2}(T)$ of P-doped materials in which measurements so far were limited to a temperature range close to zero-field $T_c$ \cite{GhoHc2BaP,BaPHc2,Welp}, and in view of suggested nodal gap structure it is of interest to get insight into the temperature dependent upper critical field of this system. 

In this article we report measurements of the anisotropic upper critical field in single crystals of optimally doped SrP122, $x$ = 0.35, using the pulsed magnetic field facility at Los Alamos National High Magnetic Field Laboratory. We find a clear difference between the temperature-dependent $H_{c2}$ for fields along and transverse to the tetragonal $c$-axis. The dependence for $H \parallel c$ is close to $T$-linear, which is very similar to the behavior found in previous studies on nodal KFe$_2$As$_2$ \cite{K122Terashima}, but very dissimilar with that of full-gap LiFeAs \cite{KyuilHc2LiFeAs}. In addition, we find monotonic suppression of $T_c$, $H_{c2,a}$ and $H_{c2,c}$ with heavy-ion irradiation.

\section{Experimental}

\begin{figure}[tb]
\includegraphics[width=8cm]{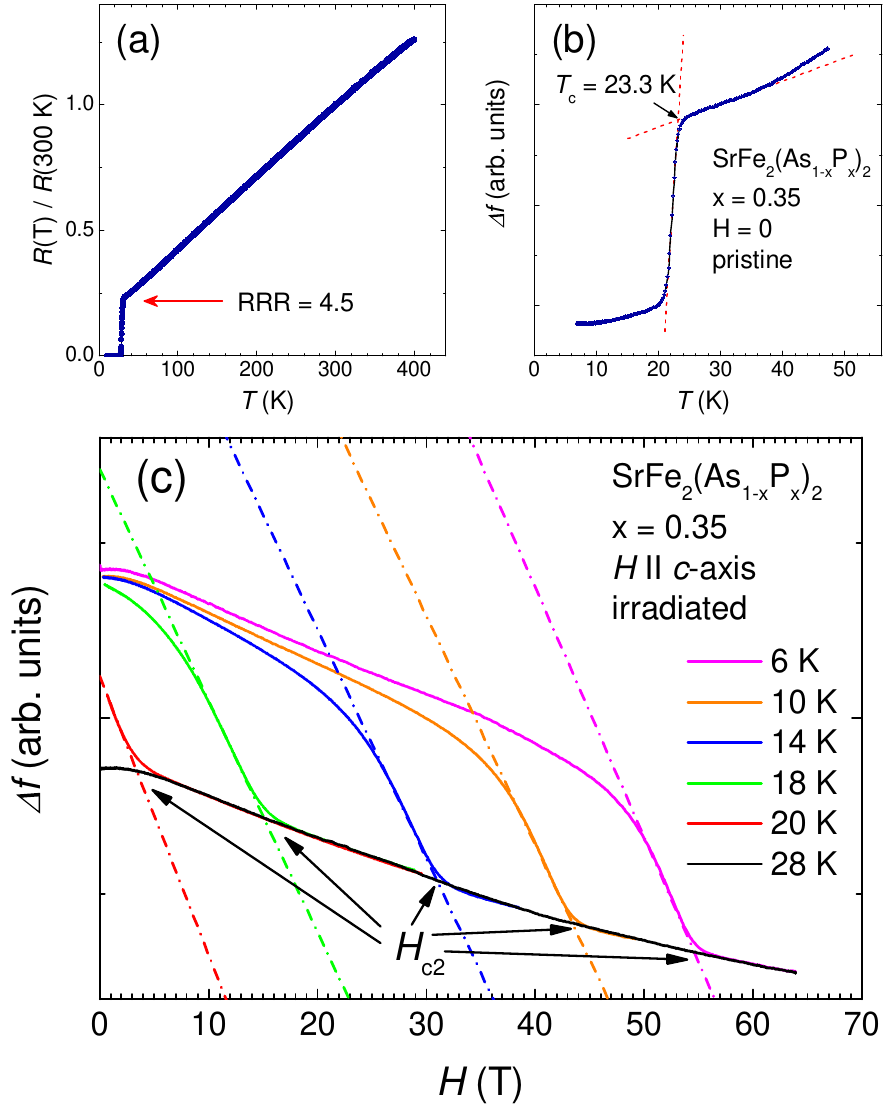}
\caption{(Color online) (a) Temperature dependent in-plane resistivity of the representative sample of pristine SrFe$_2$(As$_{1-x}$P$_x$)$_2$, $x=$ 0.35. Close to perfect $T$-linear dependence is observed in the temperature range from $T_c$ up to 400~K. Resistive transition ends at $\sim 25~K$, close to the magnetic transition observed in zero-field TDR measurements (b). The bottom panel (c) shows TDR frequency shift (in arbitrary units) measured as a function of magnetic field during pulsed field experiments at indicated temperatures. The lines show the way the $H_{c2}$ was defined from the data as a cross-over point of linear extrapolation of the rapid frequency drop to the level of the background signal. }
\label{characterization}
\end{figure}

Single-crystalline samples of SrFe$_{2}$(As$_{1-x}$P$_{x}$)$_{2}$ were grown using the self-flux method, see Refs.~\onlinecite{Kasahara,crystals} for details. Typically samples had a shape of irregular platelets, with in-plane dimensions (0.3-1)*(0.3-1) ~mm$^2$ and thickness 0.02 to 0.1 mm. A sample composition of $x$=0.35 was determined using EDX analysis. For our study we used several samples from the same batch. Four Sn-soldered contacts \cite{SUST} were attached to one of the samples to measure in-plane resistivity, plotted using normalized $\rho(T)/\rho(300K)$ in Fig.~\ref{characterization}(a). The value of $\rho(300K)$ was about 300 $\mu \Omega\cdot$cm, close to a value found in BaP122 \cite{BaPinterplane}. Samples show very close to perfect $T$-linear temperature dependence, which is similar to the dependence found in BaP122 at optimal doping, both for in-plane \cite{Kasahara} and inter-plane \cite{BaPinterplane} transport. The resistive transition is rather sharp, $\Delta T_c \sim$ 2~K, and its end point at 25~K is close to an onset of a sharp frequency shift in zero-field TDR experiments.

For this study we used two small single crystals from the same batch. One sample was irradiated with 1.4 GeV $^{208}$Pb$^{56+}$ ions at the Argonne Tandem Linear Accelerator System (ATLAS) with ion flux of ~5x10$^{11}$ ions$\cdot$s$^{-1}\cdot$m$^{-2}$.  The thickness of this sample of approximately 20 $\mu$m was much smaller than stopping distance of the ions, $\approx$ 60 $\mu$m, thus allowing creation of columnar defects. Assuming that each defect acts as a pinning center for one magnetic flux quantum, this irradiation corresponds to a matching magnetic field of approximately 25~T, corresponding to a dose of 1.2$\times$10$^{12}$ ions$\cdot$cm$^{-2}$.

Zero-field measurements on the samples used in pulsed field studies in Los Alamos NHMFL were performed in a separate TDR setup in Ames laboratory, see Ref.~\onlinecite{Prozorov2000} for details of the TDR measurements. A typical temperature-scan for pristine sample of SrP122 is shown in Fig.~\ref{characterization}(b). Onset of superconductivity at $T_c$ = 25~K provides a clear change of the tunnel diode resonance frequency. Similar frequency variations are found in isothermal pulsed magnetic field sweeps, as shown in Fig.~\ref{characterization}(c). 
%
%
During pulsed field experiments samples were glued with GE varnish to a pancake coil for the TDR setup. The whole assembly was aligned under the microscope with respect to sample holder, providing the accuracy within 2$^{\circ}$ with respect to the principal crystallographic directions.

\section{Results}

\begin{figure}[tb]
\includegraphics[width=8cm]{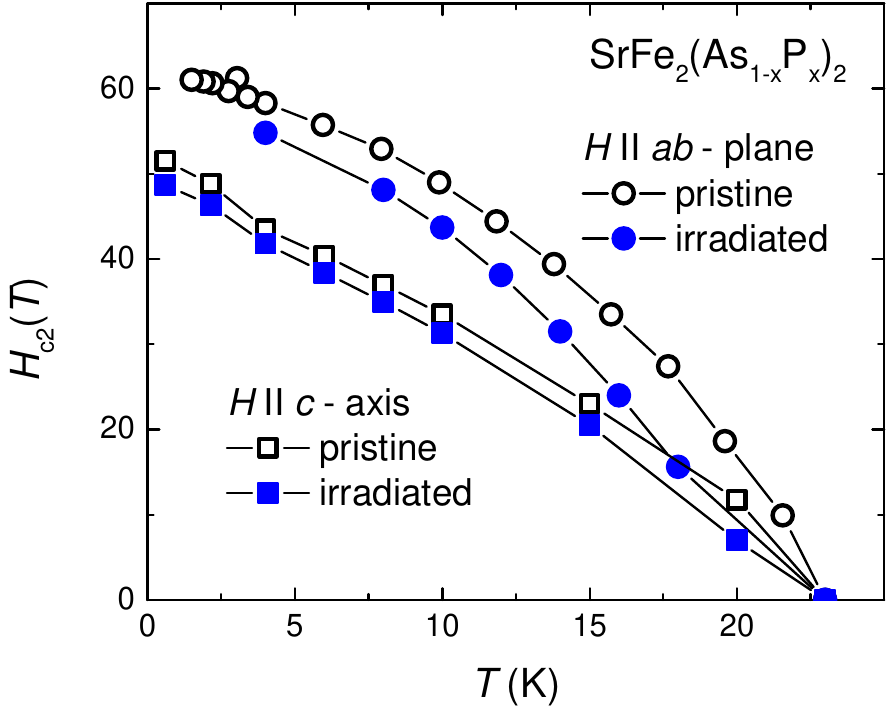}
\caption{(Color online) The temperature dependent upper critical fields for pristine (black open symbols) and irradiated (closed blue symbols) samples of optimally doped SrFe$_2$(As$_{1-x}$P$_x$)$_2$ with $x$ = 0.35, as determined from pulsed field measurements in magnetic fields parallel to the $c$-axis (squares) and to the conducting $ab-$plane (circles). The lines connect data points and do not show true position of zero-field $T_c$, determined in separate experiments. }
\label{Hc2pulsed}
\end{figure}

In Fig.~\ref{Hc2pulsed} we plot $H_{c2}(T)$ as determined from pulsed field TDR measurements at differemt temperatures for pristine and irradiated samples. Note, that zero-field $T_c$ was determined in a separate experiment, so that the $H_{c2}(T)$ behavior close to zero field $T_c(H=0)$ is not well defined. Several features of the data should be mentioned. The data for magnetic fields perpendicular to the plane field orientation, $H_{c2,c}(T)$, in general follow a $T$-linear dependence. The data for magnetic fields parallel to the plane, $H_{c2,a}$(T), show a clear tendency for saturation approaching $T=0$, which is a common expectation for both orbital \cite{WHH} and paramagnetic \cite{CC} limiting mechanisms of the upper critical field. 

\begin{figure}[tb]
\includegraphics[width=8cm]{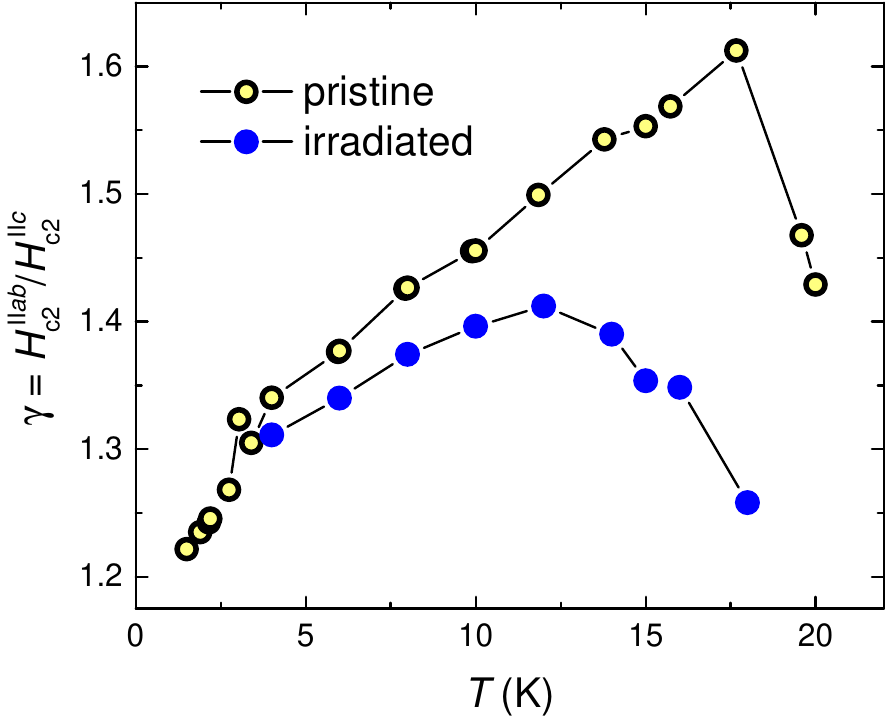}
\caption{(Color online) The temperature-dependent anisotropy parameter $\gamma(T) \equiv H_{c2,a}(T)/H_{c2,c}(T)$ in pristine (open circles) and irradiated (closed circles) samples of SrFe$_2$(As$_{1-x}$P$_x$)$_2$, $x=$ 0.35.} 
\label{gammaSrP122}
\end{figure}

In Fig.~\ref{gammaSrP122} we plot the temperature-dependent anisotropy parameter, $\gamma \equiv H_{c2,a}/H_{c2,c}$, for pristine and irradiated samples of SrP122. As is common for iron-arsenide superconductors, the anisotropy parameter is strongly temperature dependent. In most of the iron-arsenides the anisotropy parameter monotonically decreases upon cooling \cite{anisotropy,KoganProzorov,Gurevich}. 
Upon irradiation the $H_{c2}$ anisotropy decreases, whereas $T_c$ is hardly changed.  These findings are similar to the results on heavy-ion irradiated optimally doped BaK122 \cite{fang2012}. Taken in conjunction with flattening of $H_{c2,a}(T)$, this fact is frequently discussed as a signature of paramagnetic effects in magnetic fields parallel to the plane \cite{CC}. Alternatively, the same feature of monotonically decreasing upon cooling $\gamma(T)$ is explained as a consequence of strong multi-band effects in iron pnictides \cite{KoganProzorov}.

\section{Discussion}

For the description of the upper critical field in SrP122, it is important to understand the superconducting gap structure of this material. Although the detailed measurements as a function of doping were not performed yet, the available NMR \cite{SrPNMR}, microwave \cite{SrPmicrowave}, and radio-frequency London penetration depth \cite{SrPTDR} measurements suggest that the superconducting gap of optimally doped SrP122 is nodal. 

The effect of the nodal gap structure on the temperature and angular dependence of the upper critical field is a long standing problem in the field of unconventional superconductivity, stemming back to the early 80's \cite{Gork-heavy}. While these effects in multi-band superconductors may be far too complicated \cite{KoganProzorov}, we decided to take a purely empirical approach and compare the temperature dependent $H_{c2}$ for various iron-based materials. Stoichiometric superconductors provide a good reference point. There is ample evidence that the superconducting gap of LiFeAs, believed to be representative of slightly overdoped regime \cite{LiFeAsoverdoped}, is full and practically isotropic with only minor multi-band effects \cite{HKimLiFeAs,TanatarLiFeAs}. The superconducting gap of KFe$_2$As$_2$ is nodal, as suggested by a plethora of various experiments, although the exact origin (accidental $S^{\pm}$ or symmetry imposed $d$-wave) and location of the nodes is heavily debated \cite{ReidK,MatsudaKScienceARPES}. 

\begin{figure}[tb]
\includegraphics[width=8cm]{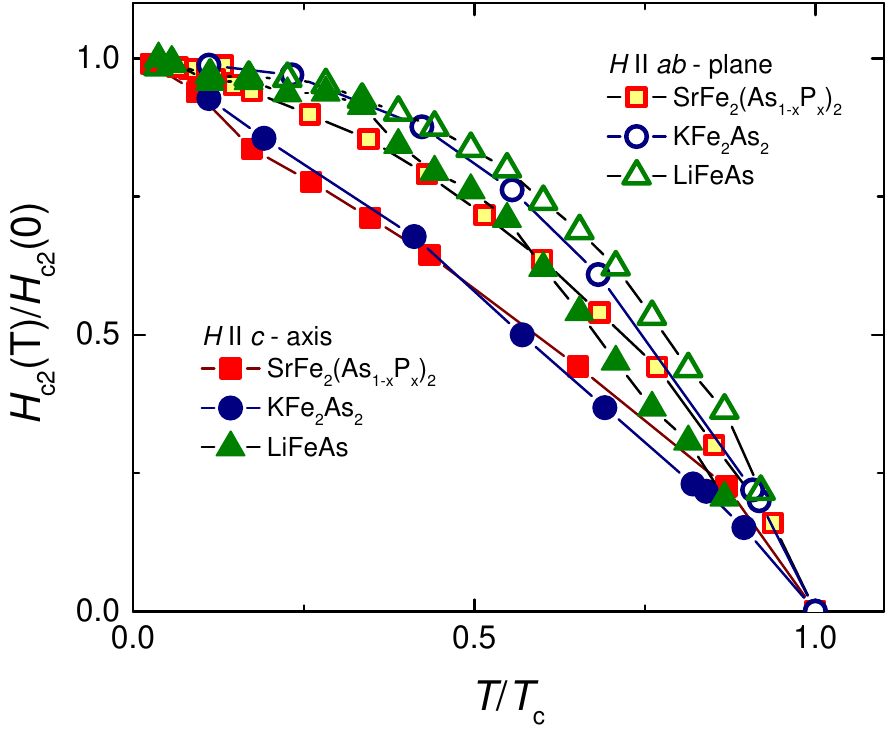}
\caption{(Color online) The $H-T$ phase diagram of SrFe$_2$(As$_{1-x}$P$_x$)$_2$, $x=$ 0.35 (red). Normalized $H/H_{c2}(0)$ is plotted vs. normalized temperature, $T/T_c(0)$. Solid symbols show $H_{c2} \parallel c-$axis and open symbols show $H_{c2} \parallel ab-$plane.  
For reference we show similar plots for full-gap superconductor LiFeAs (green up-triangles)~\cite{KyuilHc2LiFeAs}, and nodal KFe$_2$As$_2$ (blue circles)~\cite{K122Terashima}.} 
\label{Hc2vsLiK}
\end{figure}

In Fig.~\ref{Hc2vsLiK} we compare the $H_{c2}(T)$ curves for two principal directions of the applied magnetic field in SrP122 with those for LiFeAs and KFe$_2$As$_2$. One feature of the data is obvious. While the $H_{c2,c}(T)$ flattens in LiFeAs, as expected for $s$-wave superconductors in WHH theory \cite{WHH}, the dependence is roughly linear in SrP122 and KFe$_2$As$_2$, - the materials with the nodal superconducting gap. The behavior in magnetic field parallel to the plane is not as different, with all compounds showing clear signatures of saturation on approaching $T =0$. 

\begin{figure}[tb]
\includegraphics[width=8cm]{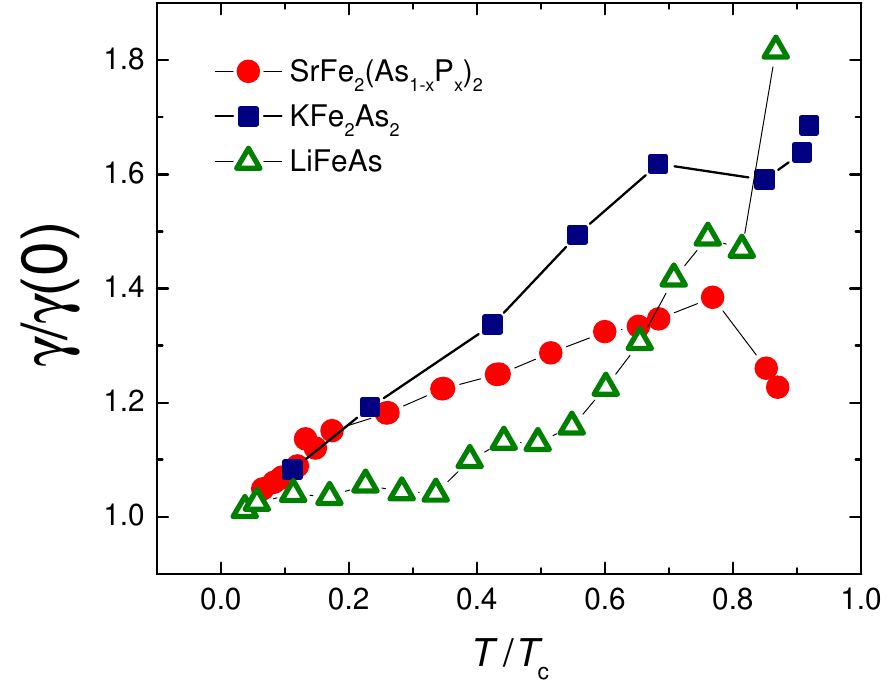}
\caption{(Color online) The temperature-dependent anisotropy parameter $\gamma(T) \equiv H_{c2,a}(T)/H_{c2,c}(T)$ normalized to its value at $T=0$ in pristine SrFe$_2$(As$_{1-x}$P$_x$)$_2$, $x=$ 0.35 (filled red circles). For the reference we show $\gamma(T/T_c)/\gamma(0)$ in clean iron-pnictide superconductors, - nodal KFe$_2$As$_2$ (blue squares)~\cite{K122Terashima}, and full-gap LiFeAs (green triangles)~\cite{KyuilHc2LiFeAs}.} 
\label{gammavsLiK}
\end{figure}

In Fig.~\ref{gammavsLiK} we compare the temperature-dependent anisotropy parameter, $\gamma(T)$, of pristine samples of optimally doped SrP122 with $\gamma(T/T_c)/\gamma(0)$ in nodal KFe$_2$As$_2$ and full-gap LiFeAs. Reflecting mainly the difference in the behavior of $H_{c2,c}(T)$, the behavior of $\gamma(T)$ is dramatically different for nodeless and nodal materials. 


While the saturation of $H_{c,a2}(T)$ on approaching $T$=0 in magnetic field parallel to Fe-As planes is in line with the predictions of both orbital and paramagnetic mechanisms of $H_{c2,a}(T)$ \cite{WHH,CC}, a near $T$-linear dependence for $H \parallel c$ is quite exotic, and its origin is not well understood. 
Several experimental studies reported nearly $T$-linear $H_{c2,c}(T)$. 
Nearly straight $H_{c2,c}(T)$ for $H \parallel c$ was observed in MgB$_2$ \cite{GurevichMgB2} and was explained in the orbital-limiting model for two-band superconductivity in the dirty limit. If the two bands are characterized by the equal diffusivities $D_1$ = $D_2$, the $H_{c2}(T)$ follows WHH type dependence \cite{WHH}, however, if the diffusivity in a weaker band $D_2$ is much smaller than the diffusivity $D_1$ in a stronger band, $H_{c2}(T)$ has upward curvature. Thus, there is a ratio $D_2/D_1$ at which $H_{c2}($T) becomes nearly straight. 
Similarly, near $T$-linear $H_{c2}(T)$ was observed in doped iron-based superconductors, BaCo122 \cite{FBS1}, BaK122, and FeSeTe \cite{FBS2}. However, due to very high values of $H_{c2}(0)$ it is not clear whether this linear behavior will change to a saturation at the lowest temperatures. In all these materials the linear $T-$dependence was explained in a similar multi-band scenario in the dirty limit, as in MgB$_2$, but the question of whether the dirty limit is achieved is still open \cite{Gurevich}. The $T-$linear dependence of $H_{c2}$ was found in other superconductors. In KOs$_2$O$_6$ it was explained in terms of orbital limiting mechanism, due to missing spatial inversion symmetry \cite{Shibauchi}. In organic superconductor $\kappa$-(BEDT-TTF)$_2$Cu[N(CN)$_2$]Br, $T$-linear dependence was found in magnetic fields parallel to the two-dimensional plane \cite{Kamiya,SM133}. Interestingly, here the behavior at ambient pressure closely follows a square root dependence as expected for paramagnetic Pauli limiting \cite{Kovalev}, while with pressure this dependence becomes close to $T$-linear \cite{Kamiya,SM133}. A similar trend is found in other organic superconductors \cite{Ishiguro}, and possible relation of $T$-linear dependence to the formation of inhomogeneous Fulde-Ferelle-Larkin-Ovchinnikov state \cite{FF,LO} was suggested by the experiments on samples with varying disorder \cite{TanatarlambdaBETS}.

\section{Conclusions}

In summary, using pulsed field tunnel-diode resonator (TDR) measurements, we have determined the upper critical fields along two principal crystallographic directions in single crystals of isoelectron-substituted SrFe$_2$(As$_{1-x}$P$_x$)$_2$ at optimal doping, $x=$ 0.35. We found that the shape of $H_{c2}(T)$ curves is different for the fields perpendicular and parallel to the tetragonal $c-$axis. It shows WHH-like behavior with clear saturation as $T \to 0$ for $H \parallel ab-$plane and practically $T$-linear variation for $H\parallel$ c-axis. We show that the shape of the $H_{c2}(T)$ curves is not affected much by heavy ion irradiation, which suppresses $T_c$ and $H_{c2}(0)$ by the same amount. We do not see any special features in the $H_{c2,c}(T)$ line, corresponding to a matching field of 25 T in heavy-ion irradiated samples. The temperature dependence of $H_{c2}$ for two principal directions in SrFe$_2$(As$_{1-x}$P$_x$)$_2$ is similar to that found in a nodal superconductor KFe$_2$As$_2$ \cite{K122Terashima}, but is different from that of fully gapped LiFeAs\cite{KyuilHc2LiFeAs}. This similarity may be suggestive that the anomalous linear shape of $H_{c2,c}(T)$ in iron pnictides may be related to a nodal superconducting gap.

\section{Acknowledgments}
We thank A.~Gurevich for useful comments. Work at the Ames Laboratory was supported by the Department of Energy-Basic Energy Sciences under Contract No. DE-AC02-07CH11358. Work at Argonne National Laboratory was supported by the U. S. Department of Energy, Office of Science, Office of Basic Energy Sciences under Contract No. DE-AC02-06CH11357. The heavy-ion irradiation was performed at the ATLAS facility at
Argonne. Work at LANL was performed under the auspices of the National Science Foundation, the Department of Energy, and the State of Florida. Work at Osaka University was supported by JST IRON-SEA.

\end{document}